\documentclass[fleqn,12pt,twoside]{article}
\usepackage{espcrc1}

\usepackage{graphicx}
\usepackage[figuresright]{rotating}

\title{Short-lived {\it p}-nuclides in the early solar system and\\
implications on the nucleosynthetic role of X-ray binaries} 

\author{
N. Dauphas\address[EFI]{Enrico Fermi Institute, The University of
Chicago, 5640 South Ellis Avenue, Chicago IL 60637, USA (dauphas@uchicago.edu)}, 
T. Rauscher\address[Basel]{Departement f{\"u}r Physik und Astronomie, Universit{\"a}t Basel, Klingelbergstrasse 82, CH-4056 Basel, Schweiz}, 
B. Marty\address[CRPG]{Centre de Recherches P\'etrographiques et G\'eochimiques, CNRS UPR 2300, 15 rue Notre-Dame des Pauvres, BP 20, 54501 Vand\oe uvre-l\`es-Nancy Cedex, France.},
and        
L. Reisberg\addressmark[CRPG]
}
       
\begin{document}

\maketitle

\begin{abstract}
\noindent
The data available for short-lived $p$-nuclides are used in an open nonlinear
model of the chemical evolution of the Galaxy in order to discuss the origin
of extinct radionuclides, the stellar sources of $p$-nuclides, and the
chronology of solar system formation. It is concluded that the observed
abundances of $^{97}$Tc, $^{98}$Tc, $^{92}$Nb, and $^{146}$Sm in the early
solar system are consistent with nucleosynthesis in type II supernovae during
continuous chemical evolution of the Galaxy and a subsequent short isolation
of the presolar molecular cloud from fresh nucleosynthetic inputs. However,
further work on supernova models is needed before
$p$-radionuclides will comprise reliable cosmochronometers. Despite these
limitations, we argue that niobium-92 can be used to test whether the {\it
rp}-process contributed to the synthesis of light {\it p}-nuclides in the
Mo-Ru region.
\end{abstract}

\section{Introduction}

Extinct nuclides are radioactive nuclides that were present in the early solar
system but have now decayed below detection levels. Their prior presence can
be inferred from abundance variations of their daughter nuclides in solar
system formation remnants. The origin of short-lived nuclides in the early
solar system is the subject of much debate. They may have resulted from ($i$)
inheritence from the chemical evolution of the Galaxy
\cite{schramm70,clayton85}, ($ii$) injection from a nearby giant star that
might have triggered the protosolar nebula into collapse \cite{cameron96},
and ($iii$) irradiation within the solar system by intense stellar flares
\cite{lee98,mckeegan00,gounelle01,chaussidon02,leya02}. In order to
investigate the formation of the solar system, one can proceed either
forwards or backwards in time. The abundances of short-lived $p$-nuclides in
the interstellar medium at solar system birth can be derived from modeling of
galactic chemical evolution \cite{schramm70,clayton85} and stellar
nucleosynthesis \cite{howard91,rayet95}. The abundances of short-lived
$p$-nuclides in the early solar system can also be derived from laboratory
measurements of solar system formation remnants. Comparison between the
predicted and the observed abundances provides unequalled information on the
origin of extinct radionuclides and solar system birth. 

\section{Abundances in the early solar system}

Four short-lived $p$-nuclides might have been alive in the early solar system;
$^{97}$Tc which EC-decays to $^{97}$Mo (mean life of 3.8 Ma), $^{98}$Tc which
$\beta$-decays to $^{98}$Ru (6.1 Ma), $^{92}$Nb which EC-decays to $^{92}$Zr
(50.1 Ma), and $^{146}$Sm which $\alpha$-decays to $^{142}$Nd (149 Ma). Among
these, only upper limits could be derived for $^{97}$Tc
\cite{dauphas01b,dauphas02c} and $^{98}$Tc \cite{becker01}, while the
abundances of $^{92}$Nb \cite{harper96a,schonbachler01} and $^{146}$Sm
\cite{lugmair92,prinzhofer92,nyquist94} are firmly established. The initial
abundances normalized to a stable neighbor nuclide synthesized by the same
process are compiled in Table 1.

\begin{table}[t]
\caption{\footnotesize Abundances of short-lived $p$-nuclides. R is the initial ratio in the
solar system derived from meteorite measurements, P is the production ratio
in supernovae, $\Re_{\rm ESS}$ is the remainder ratio in the early solar
system calculated as R/P, and $\Re _{\rm ISM}$ is the remainder ratio in the
interstellar medium derived from GCE modeling (see text for details and
references).}
{\scriptsize
\begin{tabular}{lccccccc}
\hline
\hline
Nuclide  &  & R & P & & $\Re_{\rm ESS}$  & $\Re_{\rm
ISM}$\\
\hline
Technetium-97 & ${\rm ^{97}Tc/^{98}Ru}$ & $<4\times 10^{-4}$ & $4.4\pm
1.3\times 10^{-2}$ & & $<1.3\times 10^{-2}$ & $1.2\pm 0.3\times 10^{-3}$ \\
Technetium-98 & ${\rm ^{98}Tc/^{98}Ru}$ & $<8\times 10^{-5}$ & $7.3\pm
2.5\times 10^{-3}$ & & $<1.7\times 10^{-2}$ & $1.9\pm 0.5\times 10^{-3}$ \\
Niobium-92 & ${\rm ^{92}Nb/^{92}Mo}$ & $2.8\pm 0.5 \times 10^{-5}$ &
$1.5\pm 0.6\times 10^{-3}$ & & $1.9\pm 0.8\times 10^{-2}$ & $1.6\pm 0.4\times
10^{-2}$ \\
Samarium-146 & ${\rm ^{146}Sm/^{144}Sm}$ & $7.6\pm 1.3 \times 10^{-3}$ &
$1.8\pm 0.6\times 10^{-1}$ & & $4.2\pm 1.6\times 10^{-2}$ & $4.6\pm 1.1\times
10^{-2}$ \\
\hline
\end{tabular}
}
\end{table}

\section{Stellar production ratios}
The origin of $p$-nuclides still poses some puzzles. The currently most
favored production process is by photodisintegration of seed nuclides, this is
called the $\gamma$-process. However, as discussed below, there still remains
the problem of finding the proper site for the $\gamma$-process. Furthermore,
it is unknown whether the $\gamma$-process is the only type of process that
contributed to $p$-nuclide production. In the following we discuss two 
production
sites, type II and type Ia supernovae. The feasibility of an additional
scenario, the $rp$-process, will be discussed in Sec.\ \ref{rpproc}.

\subsection{Type II supernovae}
Currently, core-collapse supernovae of type II (SNII) are strongly favored to
be the site for the so-called $p$- or $\gamma$-process outlined above. The
necessary temperatures of about $2\leq T_9 \leq 3$ can be reached during
hydrostatic and explosive neon/oxygen-burning
\cite{Woo78,Pra90,Ray90,lambert92,rayet95,Wal97,Arn99,Lan99}. However, modern
models still fail to reproduce the full range of $p$-nuclides although they
succeed in producing substantial amounts in several mass ranges. More
pronounced is the underproduction in the Mo-Ru region. For the initial
production ratios ejected from the site of the $p$-process and used in our
chemical evolution model described in Sec. 4, we have taken ratios from a
recent study of nucleosynthesis in SNII \cite{Rau02}. While $p$-nuclides with
$A>100-110$ are produced in amounts in rough  agreement with solar
abundances, for the Mo-Ru region the new model also  exhibits underproduction
features similar to those found in previous studies  \cite{Woo78,rayet95}.
The reasons for this remain unclear. Details of the production are strongly
dependent on stellar physics, such as the convection treatment used.
Furthermore, the majority of the photodisintegration rates acting in this
region are derived from theory. However, the experimental neutron capture
data \cite{bao00} as well as the rare data on proton capture in this region
\cite{Har01} agree well with the predictions \cite{rauadndt,rauadndt1}.
Additionally, an increased amount of seed nuclei would be needed in order to
account for the missing abundances \cite{Woo78}. For our purposes, the
absolute production factors are not important as we use abundance ratios.
Obviously, in the following we have to assume  that those ratios will not
change when the reason for the  underproduction is identified. We used the
SNII yields for three progenitor masses, 15, 21, and 25 M$_{\odot}$. These
yields were integrated over the initial mass function IMF, describing the
birth rate $\phi$ of stars with different initial mass $m$ ($\phi \propto
{\rm d}N/{\rm d}m$). For stars with initial mass higher than M$_{\odot}$, the
initial mass function follows a power law of the mass $\phi \propto m
^{-\alpha}$, where $\alpha=2.3\pm 0.7$ \cite{kroupa01}. In order to calculate
the IMF-averaged yields, the mass interval 12.5-27.5 was further divided into
three intervals 12.5-17.5, 17.5-22.5, and 22.5-27.5 M$_{\odot}$. The yields
of the 15, 21, and 25 M$_{\odot}$ mass stars were then weighted by the IMF
integrated over the three mass intervals. The IMF-integrated SNII production
ratios are compiled in Table\ 1.

\subsection{Type Ia supernovae}

Additional sites for the $p$-process have mainly drawn interest because of the
problem with producing the Mo-Ru region. It has been proposed that the
temperatures required for the photodisintegrations of the $p$-process could
be reached in the C-rich zone of type Ia supernovae (SNIa) \cite{howard91}.
Early claims that the problem in the Mo-Ru region could also be alleviated
have not been substantiated by making use of more realistic astrophysical
models \cite{How93,rayet95}. Howard and Meyer \cite{How93} find that they
cannot produce $p$-nuclei in sufficient quantities to make a significant
galactic contribution unless the $s$-process seeds are enhanced by  several
orders of magnitude. Furthermore, they observe that the production trends
with respect to mass number are similar to those observed in SNII models.
This can be understood by the fact that the burning during accretion on the
C/O core proceeds in a similar manner as the late burning stages of SNII
progenitors. The actual photodisintegration process is the same. The above
results seem to support the view that either SNIa do not contribute to the
$p$-ratios or, if they do, the ratios are not significantly changed from
those obtained with SNII models.

\section{Open nonlinear galactic chemical evolution}

The chemical evolution of  the Galaxy in the solar neighborhood is governed by
several mechanisms \cite{clayton85,clayton88,pagel97}. We assume that the
surface density of the galactic disk in the solar neighborhood $\Sigma$ grew
with time from zero to its present-day value as a result of the accretion of
low-metallicity gas. The infall rate is modelled as a normal density law
$\mathcal{G}$ with standard deviation equal to the mean $\tau$, where $\tau$
is the infall time-scale \cite{chang99},
\begin{equation}
{\rm \frac{d\Sigma}{d{\it t}}=\xi \, \mathcal{G}(\tau)}.
\end{equation}
The net evolution of the gas surface density ${\rm \Sigma _g}$ is a balance
between star formation and disruption on one hand, and infall of low
metallicity gas on the other hand. The rate of star formation depends on the
gas density ($\Sigma _g$) according to Schmidt's law, 
\begin{equation}
{\rm \frac{d\Sigma _g}{d{\it t}}=-\omega \, \Sigma _g^n+\xi \,
\mathcal{G}(\tau)}.
\end{equation}
For a nuclide for which the instantaneous recycling approximation is valid,
its concentration in the interstellar medium Z evolves through time as a
result of stellar nucleosynthesis (y is the yield defined as the ratio of
newly created matter of that nucleus to newly created mass of permanent
stellar remnants), dilution by infall of low-metallicity  gas, and
radioactive decay,
\begin{equation}
{\rm \frac{dZ}{d{\it t}}=y \omega \, \Sigma _g^{n-1}-\frac{\xi\,
\mathcal{G}(\tau)}{\Sigma _g} \, (Z-Z_{\xi})-\lambda Z} \quad .
\end{equation}
Three parameters  ($\xi$, $\tau$, and $\omega$) are enough to describe the
chemical evolution of the Galaxy in the solar neighborhood and to
characterize its most important features. In order to constrain these
parameters, we proceeded like Sommer-Larsen \cite{sommer91}. The present disk
surface density, gas mass fraction, and metallicity in the solar neighborhood
are known. Knowledge of these boundary conditions allows a straightforward
derivation of the dependence of $\xi$, $\omega$, and ${\rm y_o}$ upon $\tau$.
The G-dwarf metallicity distribution is used in order to estimate $\tau$ and
all other parameters which depend on it using a minimization algorithm.

The exponent of the Schmidt power law is estimated to be $1.4\pm 0.15$
\cite{gerritsen97,kennicutt98}. The initial disk density is taken to be zero
and its present-day value $45\pm 5$ ${\rm M_{\odot}\,pc^{-2}}$
\cite{holmberg00,creze98}. The gas density is estimated to be $13\pm 4$ ${\rm
M_{\odot}\,pc^{-2}}$ \cite{holmberg00}. The metallicity at solar system birth
is ${\rm [O/H]=0}$. The metallicity of the infalling gas is assumed to have
been constant through time, ${\rm [O/H]=-1}$ \cite{sommer91,wakker99}. The
age of the Galaxy is estimated to be $13.2 \pm 1.5$ Ga \cite{chaboyer01}. The
G-dwarf metallicity distribution in the solar neighborhood is taken from
Haywood \cite{haywood01}. The model that best fits the G-dwarf metallicity
distribution (Fig.~1) corresponds to $(\xi ,\tau, \omega)=(56,4.8,0.065)$.
\begin{figure}[t]
\begin{center}
\includegraphics[width=7cm]{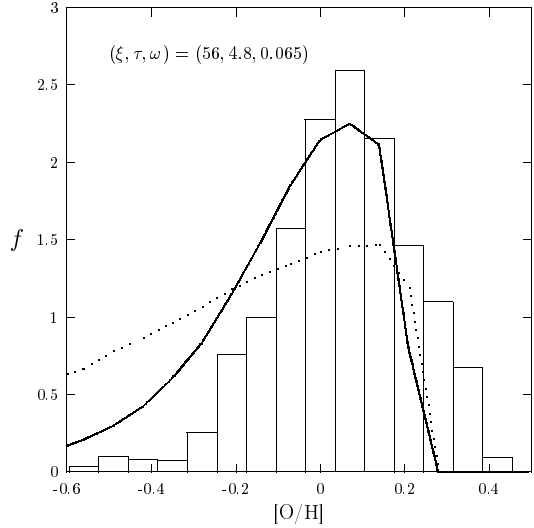}
\end{center}
\caption{\footnotesize G-dwarf metallicity distribution (density) in the solar neighborhood
\cite{haywood01}. The [Fe/H] was converted to [O/H] using ${\rm [O/H]=0.70\pm
0.04[Fe/H]+0.07\pm 0.02}$ \cite{chen00}. Histogram: observations
\cite{haywood01}, dotted line: closed-box model, solid line: best infall
model. The parameters describing the best infall model are indicated at the
top left corner.}
\end{figure}

Simulations provide a reliable means of estimating uncertainties on model
parameters. Several sets of possible observations are first simulated. Then,
each simulated set is used to compute a set of possible model parameters.
Finally, this set of possible model parameters is examined for its
distribution. The exponent of the Schmidt power law, the present surface
density of the Galaxy in the solar neighborhood, and the gas density were
simulated as normal density laws. Residuals between the best fit G-dwarf
metallicity distribution and the observed distribution were calculated. These
residuals were then resampled with repetition and added to the best fit
distribution. This bootstrap procedure permitted the simulation of a set of
possible G-dwarf metallicity distributions. Using the fitted parameters, it
is then possible to calculate the theoretical abundance of a $p$-nuclide in
the interstellar medium at the time of solar system birth and its associated
uncertainty.

\section{The remainder ratio}
The remainder ratio ($\Re$) is defined as the ratio of the abundance of a
radioactive nuclide to the abundance it would have if it were stable
\cite{clayton88}. The remainder ratio in the interstellar medium at the time
of isolation of the presolar molecular cloud from fresh nucleosynthetic
inputs ${\rm \Re_{ISM}}$ can be evaluated within the framework of the
galactic 
chemical evolution model that we have presented, 
\begin{equation}
{\rm \Re_{ISM}={\int_{\it T_{\odot}}\!\!\! \Sigma _g^{n-1} \, e^{\,
\theta+\lambda {\it t}}\, d{\it t}}\,\,/\,{\int_{\it T_{\odot}}\!\!\! \Sigma
_g^{n-1} \, e^{\, \theta+\lambda {\it T}_{\odot}}\, d{\it t}}},
\end{equation}
where $\theta$ is the cycle number defined as ${\rm d \theta /d{\it t}=\xi
\mathcal{G}(\tau)/\Sigma _g}$ \cite{clayton88}. Using the parameters
calculated in the previous section, it can be shown that for short-lived
nuclides, the remainder ratio takes the form, 
\begin{equation}
{\rm \Re_{ISM}=\kappa /(\lambda T_{\odot})},
\end{equation}
 with $\kappa=2.7\pm 0.4$, where the confidence interval accounts for model
uncertainties and ${\rm 
T_{\odot}}$ is the presolar age of the Galaxy $8.7\pm 1.5$ Ga, $\kappa$=1
corresponds to the closed-box model. Clayton \cite{clayton85} estimated
within a linear model that $\kappa$ must be within 2 and 4. The remainder
ratio within the solar system at solar system birth ${\rm \Re_{ESS}}$ can
also be estimated using stellar production ratios and observed abundances in
meteorites,
\begin{equation}
{\rm \Re_{ESS}=R/P},
\end{equation}
where R is the early solar system ratio of a radioactive nuclide to a stable
nuclide produced by the same process and P is the production ratio. There
might have been a delay between isolation of the presolar molecular cloud
from fresh nucleosynthetic inputs and solar system formation,
\begin{equation}
{\rm \Re_{ESS}=\Re_{ISM}\,e^{-\lambda \Delta}},
\end{equation}
where $\lambda$ is the decay constant and $\Delta$ is the free decay 
interval. If all short-lived $p$-nuclides were derived from the chemical
evolution of the Galaxy, they should define a unique free decay interval. It
is worthwhile to note that the deterministic chemical evolution model of the
Galaxy probably breaks down for very short-lived nuclides because granularity
might play a role, which would require a stochastic treatment \cite{meyer97}.
The remainder ratios of ${\rm ^{97}Tc}$, ${\rm ^{98}Tc}$, ${\rm ^{92}Nb}$,
and ${\rm ^{146}Sm}$ are compiled in Table 1. As can be seen in Fig. 2, they
are consistent with synthesis of short-lived $p$-nuclides in supernovae and
derivation from the interstellar medium with a comparatively short free decay
interval ($\Delta  <100$ Ma).

\begin{figure}[t]
\begin{center}
\includegraphics[width=7cm]{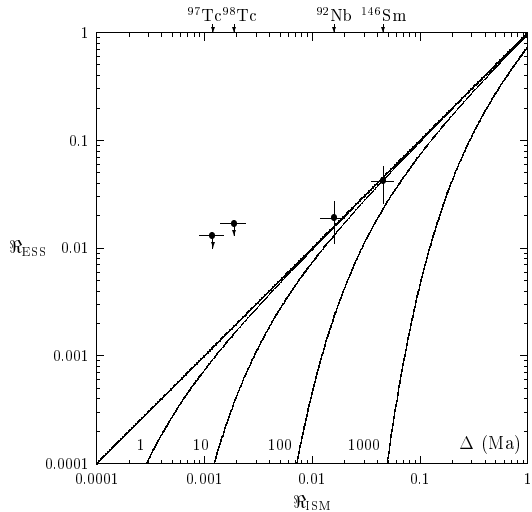}
\end{center}
\caption{\footnotesize The remainder ratio in the early solar system $\Re _{\rm ESS}$ is
shown as a function of the remainder ratio in the interstellar medium at
solar system birth $\Re _{\rm ISM}$ for a variety of free decay intervals
$\Delta$. The points represent the positions of short-lived $p$-nuclides in
this diagram (Table 1, see text for details).} 
\end{figure}

\section{Origin of extinct radioactivities}
\label{sec:bridge}

Assuming that there was no free decay  interval between isolation of the
presolar molecular cloud and condensation of the first  solids in the solar
system, our galactic chemical evolution model predicts that the initial ${\rm
^{97}Tc/^{98}Ru}$ and ${\rm ^{98}Tc/^{98}Ru}$ ratios at solar system birth
were $5.3\pm 2.0\times 10^{-5}$ and $1.4\pm 0.6\times 10^{-5}$, respectively.
These values are consistent with the upper-limits derived from measurements
(Sec.~1).

 There are a few very short-lived nuclides that cannot easily be explained by
long-term chemical evolution of the Galaxy. These short-lived nuclides are
${\rm ^{7}Be}$, ${\rm ^{10}Be}$, ${\rm ^{26}Al}$, ${\rm ^{53}Mn}$, and ${\rm
^{41}Ca}$. They may have been injected in the nascent solar system by a
nearby giant star that might have triggered the protosolar cloud into
collapse 
\cite{cameron96}. They may also have been synthesized within the solar system
by local irradiation \cite{lee98,mckeegan00,gounelle01,chaussidon02,leya02}.
Both models face difficulties but the recent discovery of ${\rm ^{10}Be}$
\cite{mckeegan00,sugiura00} and possibly ${\rm ^7 Be}$ \cite{chaussidon02} in
refractory inclusions favors the local irradiation model
\cite{lee98,mckeegan00,gounelle01,chaussidon02,leya02}.

\section{Niobium-92 and $rp$-nucleosynthesis in X-ray binaries}
\label{rpproc}

As discussed previously, supernovae models fail to reproduce the absolute
abundance of $p$-nuclides in the Mo-Ru region. This failure may be due to an
inadequacy of the stellar models, involved nuclear physics, or might 
indicate that other stellar sources
contributed to the nucleosynthesis of $p$-nuclides. Interestingly, niobium-92
is a $p$-nuclide located in the mass range that shows a severe
underproduction feature. Thus, this nuclide may provide useful constraints on
$p$-nucleosynthesis. As indicated by other extinct $r$ and $p$ nuclides, it
seems that the free decay interval was short, probably lower than a few hundreds of  million years.
Knowing the free decay interval, it is possible to do the reverse calculation and to
compute the stellar ${\rm ^{92}Nb/^{92}Mo}$ production ratio using the
observed abundance in the early solar system and the galactic chemical
evolution model presented previously. We thus estimate that the stellar ${\rm
^{92}Nb/^{92}Mo}$ ratio was necessarily higher than $1.7\pm 0.5\times 10^{-3}$ ($\Delta>0$) and was probably lower than 0.1 ($\Delta<200$ Ma).

Another means of producing proton-rich nuclei is the $rp$-process
\cite{schatz98}. In X-ray binaries, a thermonuclear runaway is ignited in the
proton-rich layer accreted on the surface of a neutron star by mass flow from
a companion star \cite{walla81,taam93}. It has been shown that proton-rich
nuclei can be synthesized in the subsequent explosive H and He burning. The
$rp$-process path closely follows the proton drip-line. In the final
freeze-out phase, the nuclides in the $rp$-path decay towards stability and
form $p$-nuclides \cite{schatz98}. Recently, a definite endpoint of the
$rp$-process was identified which confines the possible production of
$p$-nuclides to the mass region of about $A<110$ \cite{schatz01}. Thus, this
scenario would not interfere with the production in SNII but rather might
account for the missing abundances in the Mo-Ru region.  Among the isotopes
considered here, the $rp$-process could contribute to the abundances of
$^{92}$Mo, $^{97}$Tc, and
$^{98}$Ru. The nuclides $^{92}$Nb and $^{98}$Tc cannot be synthesized because
they are shielded by stable nuclides.

Niobium-92 is the only $p$-nuclide in the Mo-Ru region that cannot be 
synthetized in X-ray bursts but was alive in the early solar system.  Thus,
the abundance of this nuclide in the solar system may be used to test 
whether X-ray bursts significantly contributed to the synthesis of
proton-rich  nuclides. If X-ray bursts produced 90\% of the $p$-nuclides in
the Mo-Ru  region and supernovae synthesized the rest, then the ${\rm
^{92}Nb/^{92}Mo}$ ratio expelled from stars to the interstellar medium would
have been lowered by a factor of 10 relative to supernovae because X-ray
bursts produce no $^{92}$Nb. Thus, the apparent production ratio would have
been $1.5\pm 1.0\times 10^{-4}$, a factor of 10 lower than the required minimum
production ratio of $1.7\pm 0.5\times 10^{-3}$.  This result speaks against a
significant  contribution of X-ray binaries to the synthesis of $p$-nuclides
in the Mo-Ru region. Note that if X-ray binaries contributed to the synthesis
of $p$-nuclides in the Mo-Ru region, they would overproduce nuclides in the
A=104 region \cite{schatz01}. The problem of the origin of the Mo and Ru
$p$-nuclei therefore remains unsolved and clearly warrants further research. 


We thank F.-K. Thielemann, H. Schatz, A.M. Davis, and G.W. Lugmair for
extended discussions. This work is supported in part by the French CNES/INSU
(PNP), the NASA (grant NAGW-9510), the Swiss NSF (grant 2000-061031.02), and
the European Union (grant SMT4-CT98-2220). T. R. acknowledges support by a
PROFIL professorship (Swiss NSF grant 2024-067428.01).

\end{document}